\documentclass[conference]{IEEEtran}

\usepackage{amssymb,amsmath} 
\usepackage{graphicx}   
\usepackage{subfigure}  
\usepackage[boxed,figure,noline]{algorithm2e}   

\begin{document}

\title{Dual MCDRR Scheduler for Hybrid TDM/WDM Optical Networks} 

 \author{\IEEEauthorblockN{Mithileysh Sathiyanarayanan, \textit{IEEE Student Member} and Babangida Abubhakar, \textit{IEEE Member}}
   \IEEEauthorblockA{University of Brighton, UK\\
     Email:{m.sathiyanarayanan, b.abubakar}@brighton.ac.uk}
 }

\maketitle

\begin{abstract}
  In this paper we propose and investigate the performance of a dual multi-channel
  deficit round-robin (D-MCDRR) scheduler based on the existing single MCDRR scheduler. The existing scheduler is used for multiple channels with tunable transmitters and fixed receivers in hybrid time division multiplexing (TDM)/wavelength division multiplexing (WDM) optical networks. The proposed dual scheduler will also be used in the same optical networks. We extend the existing MCDRR scheduling algorithm for n channels to the case of considering two schedulers for the same n channels. Simulation results show that the proposed dual MCDRR (D-MCDRR) scheduler can provide better throughput when compared to the existing single MCDRR scheduler.
\end{abstract}

\begin{IEEEkeywords}
  Multi-channel scheduling, fair queueing, tunable transmitters, hybrid TDM/WDM,
  quality of service (QoS).
\end{IEEEkeywords}


\section{Introduction}

The importance of scheduling was realized several years ago for the case of single channel communication. Currently, lot of research is on multi-channel communication. The multi-channel scheduling and its applications rely on the ability of the scheduler to provide quality of service (QoS) guarantees. There are several measures that are to be considered when choosing a scheduling algorithm. The most important are fairness, latency, and complexity. In multi-channel scheduling, it is the scheduling algorithm that is key to achieve high performance. The major focus is on the delay and throughput performance of the whole system, but there is hardly any support for fairness and QoS guarantee. 

Several research works has been conducted on scheduling with different performance objectives. Dual scheduling algorithm was considered by \cite{chen:dsa}. The algorithm uses rate control and queue-length based scheduling to allocate resources for a generalized switch. In this research a new architecture was motivated by the dual scheduling algorithm in which an additional queue was introduced to interface the user data queue and to modulate the scheduling process to achieve different performance objective. This research work did not consider the quality of service guarantee. 

Another research work was conducted by \cite{wang:stochastic} where a Stochastic Primal-Dual (SPD) algorithm for downlink/uplink scheduling of multiple connections with rate requirements was derived.  In the derived algorithm, each connection transmits using adaptive modulation and coding over a wireless fading channel Based on quantized channel state information at the transmitters. In this research the authors developed a SPD algorithm which can dynamically adapt the scheduling policy when the fading statistics are not known. Also less consideration was given to the quality of service in their derived algorithm. 

A new packet scheduling algorithm for a satellite Long Term Evolution (LTE) network which adopts MIMO technology was proposed in \cite{aiyetoro:nps}. The new scheduling algorithm tagged QoS-Aware Fair (QAF) scheduler that provides a good trade-off among quality of service, fairness and throughput. However, the proposed algorithm uses single scheduler, as such it may not perform well as the amount of traffic increases. The recent developments include, introducing WDM into Next-Generation Access Networks.

The onset of wavelength division multiplexing (WDM) technology demands better packet scheduling in the multi-channel communication, especially with tunable transmitters (tunable transmitters and receivers) for hybrid time division multiplexing (TDM)/wavelength division multiplexing (WDM) systems. SUCCESS-HPON architecture in \cite{Kim:05-1} provides detailed investigation of several multi-channel scheduling algorithms under realistic environments for tunable transceivers in hybrid TDM/WDM optical networks. In this paper, we consider only tunable transmitters and fixed receivers. The main objective of this paper is to use dual multi-channel scheduler in hybrid TDM/WDM optical networks with tunable transmitters to achieve better throughput and fairness. 

Existing MCDRR scheduler for multi-channel communication \cite{mit:mcdrr} is based on the deficit round-robin (DRR) scheduling for single channel case \cite{Shreedhar:96-1}. MCDRR algorithm also considers the simple round-robin with deficit counters as in the case of DRR. Round-robin scheduling is used for servicing the queues with a quantum of service assigned to each flow. When the channel and tunable transmitter are available, scheduling is triggered and round-robin pointer starts from the first flow. The MCDRR scheduler selects packets to send out from all flows that have queued. For each flow, two variables called \textit{quantum} and \textit{deficit counter} are maintained. Quantum is a variable in bytes set to a scheduling operation dispensed to a flow for a period of one round. Deficit counter is also a variable in bytes set for each of the flow in the scheduling operation. Initially, deficit counter is set equal to the quantum. If the packet size is less than or equal to the quantum size, packet will be served. If the packet size is greater than the quantum size, packet cannot be served and it has to wait for the next round when the remainder from the previous quantum is added to the quantum for the next round. Queues that are not completely serviced in a round are compensated in the next round. The nearly perfect fairness is provided by the MCDRR has been demonstrated through simulation experiments \cite{mit:mcdrr}. 

As said earlier, dual scheduling algorithm \cite{chen:dsa} was implemented for a generalized switch where the QoS guarantee was not considered. We proposed a dual multi-channel deficit round-robin scheduler based on the existing single MCDRR scheduler that take into account the required quality of service.   

In this paper we propose and investigate the performance of a dual multi-channel deficit round-robin (D-MCDRR) scheduling algorithm, which can provide better fairness (in terms of throughput) compared to the single MCDRR scheduling algorithm. The rest of the paper is organized as follows. In Section
II we explain the concept of D-MCDRR scheduling algorithm for the multi-channel case along with the explanation of enqueuing and dequeuing processes. In Section III, we illustrate the D-MCDRR example to understand the algorithm. In Section IV, we present simulation results for the D-MCDRR algorithm by comparing the existing single MCDRR algorithm. Section V concludes our discussions in this paper. 

\section{Dual Multi-Channel Deficit Round-Robin (D-MCDRR)}

\begin{figure}[hbtp]
  \centering
  \includegraphics*[width=\linewidth]{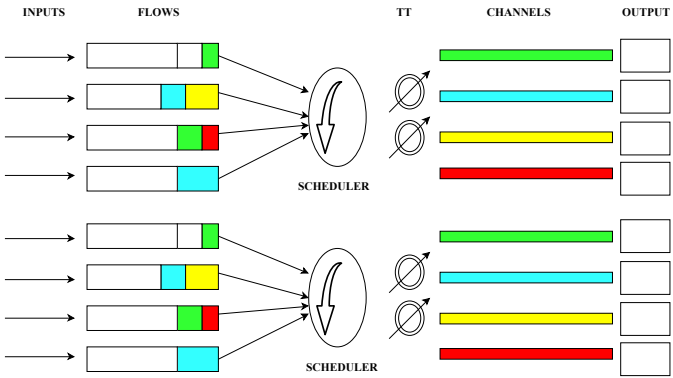}  
  \caption{Block diagram of using dual MCDRR scheduler in hybrid TDM/WDM optical networks.}
  \label{fig:mcdrr}
\end{figure}

Multi-Channel Deficit Round-Robin (MCDRR) is used in the case of multi-channel communication with tunable transmitters and fixed receivers. The dual MCDRR (D-MCDRR) is also used for the same purpose as shown in Fig. \ref{fig:mcdrr}.

The proposed D-MCDRR is an extension of the MCDRR considering: availability of the channels and tunable transmitters and overlaps `rounds' in scheduling to efficiently utilize channels and tunable transmitters. The virtual output queues (VOQs) are serviced by the simple round-robin algorithm with a quantum of service assigned to each queue as in the case of MCDRR. Dual MCDRR (D-MCDRR) scheduling algorithm is built on the existing MCDRR (single scheduler). At each round, enqueue and dequeue process takes place.

Like in MCDRR case, enqueue process is a standard operation to classify and place a packet $p$ into a VOQ for channel $i$. Dequeue process returns a round robin pointer to the head-of-line (HOL) packet in the selected VOQ or $NULL$ when the scheduler cannot find a proper packet to transmit.

For each $VOQ[i]$, we maintain the following counters as in case of single MCDRR scheduler \cite{mit:mcdrr}:
\begin{itemize}
\item $DC[i]$: It contains the byte that $VOQ[i]$ did not use in the previous round.
\item $numPktsScheduled[i]$: It counts the number of packets scheduled for transmission during the service of $VOQ[i]$. Unlike the original DRR, we nee this counter to keep track of those packets scheduled for transmission due to multiple rounds overlapped and running in parallel.
\end{itemize}
\
For D-MCDRR operation at the end of each round, the following steps take place.

 \begin{enumerate}
 \item[Step1] Initialize the pointer to the first channel and set the deficit counter to zero.
 
 \item[Step2] For n channels, dual scheduling process will take place. If the total number of channels are even (i.e. $n$ = even no. of channels), then the first scheduler takes $n/2$ channels and the other scheduler takes $n-(n/2)$ channels. If the total number of channels are odd (i.e. $n$ = odd no. of channels), then the first scheduler takes $n/2+1/2$ channels and the other scheduler takes $n-(n/2+1/2)$ channels. 
 
 \item[Step3] On the arrival of a packet $p$ from channel $n$, if Enqueue $(i,p)$ is successful, then check for tunable transmitter availability. 
 
 \item[Step4] The moment the tunable transmitter is available, it triggers the scheduling process and the first scheduler's round robin pointer starts from the Flow 1 $(VOQ[1])$ and the second scheduler's round robin pointer starts from $(VOQ[n/2])$ or $(VOQ[n/2+1/2])$ based on the total number of channels.
   
 \item[Step5] At the start of the first round, deficit counter is equal to the quantum size. After every round, deficit counter becomes equal to the previous deficit counter credits plus the quantum size.
   
 \item[Step6] If the packet size is lesser than or equal to the deficit counter size, then the packet is served. If the packet size is greater than the deficit counter size, then the packet will be served only when the deficit counter size becomes greater than packet size (in the subsequent rounds).
   
 \item[Step7] Based on the tunable transmitter and channel availability, the dequeuing process takes place and only one packet from each flow is served, then the deficit counter is updated. The deficit counter for the particular flow is reduced by the size of the transmitted packet.
   
 \item[Step8] The pointer moves sequentially for their respective channels. Once the pointer moves through all the given flows, we say it as ``Completion of One Round''. Multiple rounds overlap and run in parallel, the scheduling and the transmission of packets are not necessarily sequential as in case of single MCDRR scheduling. The whole procedure is repeated from the Step 1 through 7 until all the flows completely becomes empty.
 \end{enumerate}
\
These are the steps for D-MCDRR operation at the end of each round that takes place. So this scheduling operation depends on availability of channels, availability of data packet and availability of tunable transmitters. The following example will give better idea how the algorithm works.

\section{D-MCDRR Example}

\begin{figure*}[hbtp]  
  \centering
  \includegraphics*[width=0.46\linewidth]{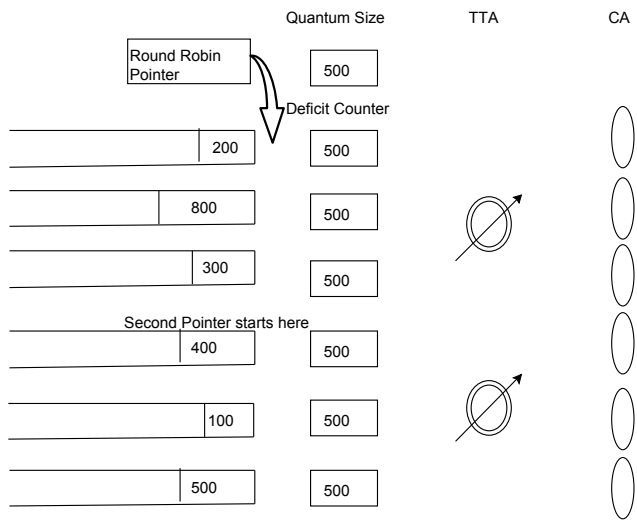}  
  \caption{Start of Round 1.}
  \label{fig:round1}
\end{figure*}

\begin{figure*}[hbtp]
  \centering
  \includegraphics*[width=0.46\linewidth]{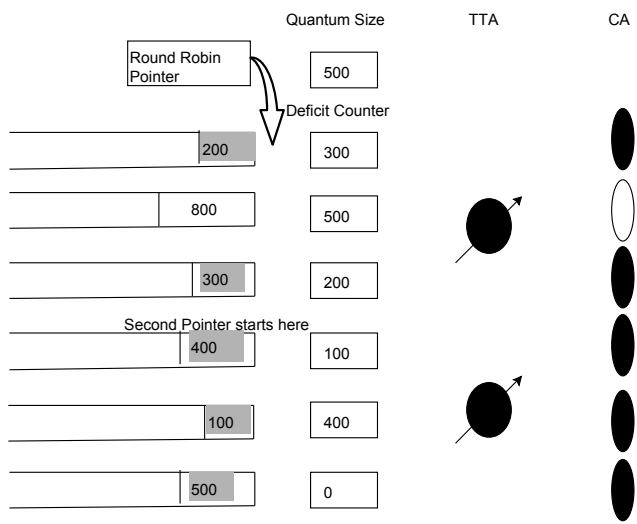}  
  \caption{End of Round 1.}
  \label{fig:round1a}
\end{figure*}

\begin{figure*}[hbtp]
  \centering
  \includegraphics*[width=0.47\linewidth]{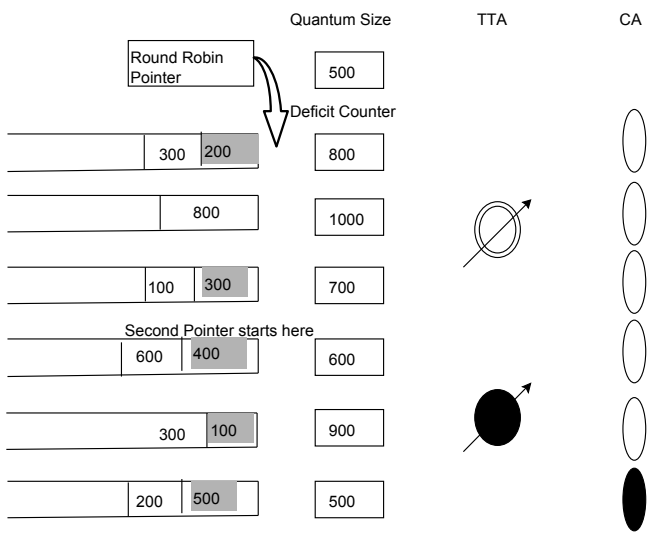}  
  \caption{Start of Round 2.}
  \label{fig:round2}
\end{figure*}

As with the MCDRR single scheduler, scheduling process starts with the triggering of the tunable transmitter which means the tunable transmitter is available to transmit the data. With the triggering, round robin pointer is activated. The two schedulers maintain two round robin pointers: one starts from the first flow and the other from the mid flow. The deficit counter gets equivalent to the quantum size. In the event that the packet size is lesser than the deficit counter and channel is accessible at that moment of time, the packet is served. On the off chance that the channel is not accessible, the pointer is moved to the following flow. At the point when the channel gets accessible then the packet will be transmitted in the following round. 

In this example, we consider 6 flows which has 6 dedicated channels along with 2 tunable transmitters and two round robin pointers are used. Quantum size is considered to be 500 credits where as deficit counters are set to zero initially. Note that, TTA and CA means tunable transmitter availability and channel availability respectively.

\textbf{Start of Round 1:}

Since the quantum size is acknowledged to be 500 credits, DC is set to 500 credits as shown in Fig. ~\ref{fig:round1}. At the start of scheduling process, both the tunable transmitters are accessible. By default, \textbf{transmitter 1} is considered and this triggers the scheduling process. The first pointer begins from the Flow 1 and simultaneously second pointer starts from the mid Flow 4. Pointer 1 starts at Flow 1 and the packet of size 200 bytes will be served since it is short of what the deficit counter 500 credits and the channel is accessible at that moment of time. By default they pick tunable transmitter 1. In the wake of serving, the deficit counter is overhauled, that is DC turns into 300 credits. Meanwhile, \textbf{transmitter 2} is available. The pointer 2 which starts at Flow 4 and the packet of size 400 bytes will be served since it is short of what the deficit counter 500 credits and the channel is accessible at that moment of time. Since tunable transmitter 1 busy serving, the scheduler picks tunable transmitter 2. In the wake of serving, the deficit counter is overhauled, that is DC turns into 100 credits.

After serving the packets, the pointer 1 moves to Flow 2 and pointer 2 moves to Flow 5. In Flow 2, packet size is greater than the deficit counter, so packet cannot be served and pointer 1 moves to Flow 3. Whereas in Flow 5, packet size is less than the deficit counter. Based on the tunable transmitter availability, packet is served. Now the pointer 2 moves to Flow 6. Both the Flow 3 and 6 have packets less than or equal to the deficit counter respectively and both the packets in respective flows are served successfully based on the availability of tunable transmitters. That is the End of Round 1 as shown in Fig. ~\ref{fig:round1a}, as both the pointers have gone through all the flows. According to the authors in the previous paper, call it as ``Completion of one Round''. 

\textbf{Start of Round 2:}

After the completion of the first round, it is the start of the second round, deficit counters are added by 500 credits as shown in Fig. ~\ref{fig:round2}. Now the transmitter 2 is available and the scheduling process continues as there are packets to be served. Similar to the first round, the first pointer begins from the Flow 1 and simultaneously second pointer starts from the mid Flow 4. Pointer 1 starts at Flow 1 and the packet of size 300 bytes will be served since it is short of what the deficit counter 500 credits and the channel is accessible at that moment of time. In the wake of serving, the deficit counter is overhauled, that is DC turns into 500 credits. Meanwhile, transmitter 2 is available. The pointer 2 which starts at Flow 4 and the packet of size 600 bytes will be served since it is equal to the deficit counter 600 credits (100+500) and the channel is accessible at that moment of time. Since tunable transmitter 1 busy serving, the scheduler picks tunable transmitter 2. In the wake of serving, the deficit counter is overhauled, that is DC turns into 0 credits.

After serving the packets, the pointer 1 moves to Flow 2 and pointer 2 moves to Flow 5. In Flow 2, packet size is now less than the deficit counter, so packet of 800 bytes can be served. Also in Flow 5, packet size is less than the deficit counter. Based on the tunable transmitter availability, packet is served. Now the pointer 1 and pointer 2 moves to Flow 3 and Flow 6 respectively. Both the Flow 3 and 6 have packets less than or equal to the deficit counter respectively and both the packets in respective flows are served successfully based on the availability of tunable transmitters. That is the End of Round 2, as both the pointers have gone through all the flows. This is another completion of round.

Since transmitter 2 is available after the completion of round 2, the next round starts from the
Flow 1 and the process continues till all the flows completely become empty which is sequential. In this way, all the packets are transmitted successfully. This example covers all the details such as packet size lesser than the deficit counter with channel available and channel not available at some instant of
time, then packet size greater than the deficit counter with channel available
and not available, flow being empty in one particular round.

\section{Simulation Results}

To demonstrate the performance of the proposed dual MCDRR (D-MCDRR) over the single MCDRR scheduling algorithm, we carried simulation experiments with a existing model for a hybrid TDM/WDM link with
tunable transmitters and fixed receivers shown in Fig. \ref{fig:mcdrr}.

\textbf{Case 1:} We set the values in the model for even number of flows as:

\begin{itemize}
\item Number of wavelengths/channels ($W$) = 20
\item Line rate of each channel = 1 Gb/s, and 
\item Number of tunable transmitters ($M$) = 2.
\end{itemize}

\begin{figure}[hbtp]
  \centering
  \includegraphics*[width=\linewidth]{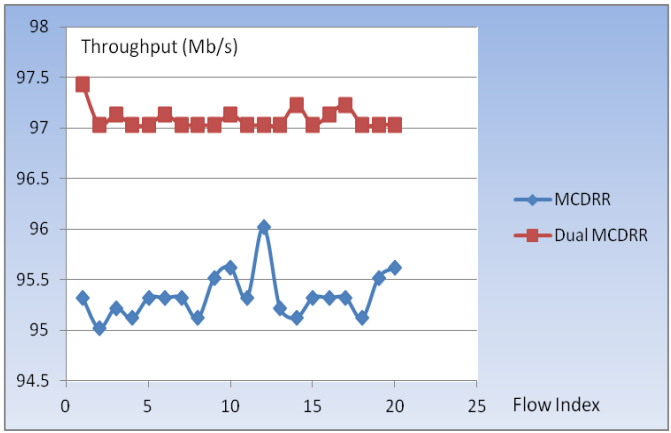}  
  \caption{Throughput for \textbf{even flows} (20 flows) with exponential inter-frame times and random frame sizes, comparing single mcdrr scheduler and the proposed dual mcdrr scheduler.}
  \label{fig:result1}
\end{figure}

\textbf{Case 2:} We set the values in the model for odd number of flows as:

\begin{itemize}
\item Number of wavelengths/channels ($W$) = 25
\item Line rate of each channel = 1 Gb/s, and 
\item Number of tunable transmitters ($M$) = 2.
\end{itemize}

\begin{figure}[hbtp]
  \centering
  \includegraphics*[width=\linewidth]{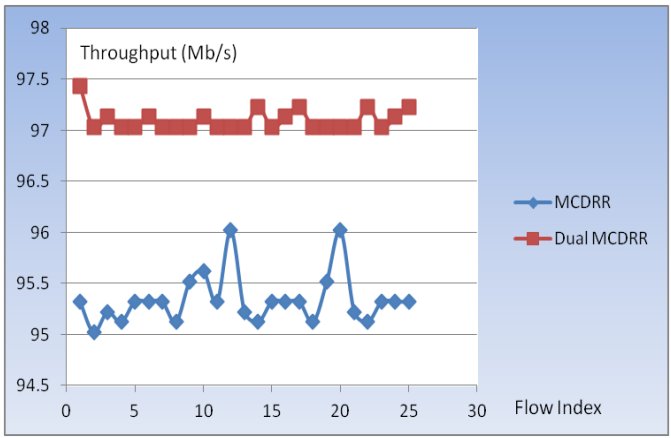}  
  \caption{Throughput for \textbf{odd flows} (25 flows) with exponential inter-frame times and random frame sizes, comparing single mcdrr scheduler and the proposed dual mcdrr scheduler.}
  \label{fig:result2}
\end{figure}

As the previous paper on MCDRR, we consider that the scheduling is done at the data link layer with
Ethernet frames and ignore the tuning time of tunable transmitters in simulation. Each VOQ can hold up to 1000 frames. We measure the throughput of each flow at a receiver for 10 mins of simulation time.

Fig. \ref{fig:result1} and \ref{fig:result2} show the throughput for even flows (20 flows) and odd flows (25 flows) with same sets of conditions based on inter-frame times and frame sizes. In both the cases, the inter-frame times are exponentially distributed with the average 48 $\mu$s for all the flows, while the frame sizes are uniformly distributed between 64 and 1518 bytes for all the flows. 

For both the cases: even and odd flows, the proposed dual MCDRR (D-MCDRR) scheduler performs much better than the single MCDRR scheduler. Raj Jain's fairness index \cite{Jain:84} for the
results are given below:

\begin{itemize}
\item In Fig. \ref{fig:result1}, single MCDRR  is 0.9999756 and dual MCDRR is 0.9999854, and
\item In Fig. \ref{fig:result2}, single MCDRR  is 0.9999798 and dual MCDRR is 0.9999876
\end{itemize}

From the simulation results, we found that the proposed dual MCDRR (D-MCDRR) scheduling
algorithm provides slightly better fairness for both even and odd flows considering conditions like inter-frame times and frame sizes compared to single MCDRR scheduling algorithm. 

\section{Conclusion}
In this paper we proposed and investigated the performance of the dual multi-channel deficit round-robin (D-MCDRR) scheduler based on the existing single MCDRR schedule with tunable transmitters and
fixed receivers in hybrid time division multiplexing (TDM)/wavelength division multiplexing (WDM) optical networks. In extending the MCDRR to D-MCDRR, we try to efficiently utilize the network resources (i.e., channels and tunable transmitters) by overlapping rounds, while maintaining its low
complexity (i.e., $O(1)$). Simulation results show that the proposed dual MCDRR (D-MCDRR) scheduler can provide better throughput and fairness when compared to the existing single MCDRR scheduler. More simulation results can be produced considering delay and latency. Performance test can be implemented considering wide range of parameters in the simulation model. The current study is on comparing different multi-channel scheduling algorithms and establishing mathematical bounds for the fairness and latency.

\section*{Acknowledgment}
We are very thankful to Dr. Kyeong Soo Kim for permitting us to use the existing MCDRR code.


\bibliographystyle{IEEEtran}
\bibliography{IEEEabrv,kks}

\end{document}